\documentclass{PoS}

\title{Recombination in nuclear collisions}

\ShortTitle{Recombination in nuclear collisions}

\author{\speaker{Rudy Hwa}\\
        Institute of Theoretical Science, University of Oregon, Eugene, OR 97403, USA\\
        E-mail: \email{hwa@uoregon.edu}}

\abstract{
Recombination is a hadronization process that converts partons to hadrons at late time, but the description has no quantitative significance without some meaningful input on the parton distributions at earlier time. Thus observations of particle spectra and correlations have definitive implications on the partonic processes at all transverse momenta. After presenting a general review of the subject at the Workshop, I selected two topics in nuclear collisions for more detailed discussion, which are summarized here. One is on the azimuthal anisotropy at low $p_T$ due to hard or semihard scattering of partons that create ridges with or without triggers. The ridge particles are the products of  recombination of thermal partons enhanced by  the energy loss of hard or semihard partons. Their $\phi$ dependence at midrapidity can be determined essentially from geometry. The other topic is on the scaling behavior in $\phi$ and centrality at high $p_T$, where the hadronization process is dominated by TS and SS recombination. The relevant RHIC data are found to have the same scaling behavior. But at LHC such scaling is badly broken at $p_T \sim 10$ GeV/c if two-jet recombination is important. At the end of this article some comments are made to relate our study of the effects of semihard partons to the observation of minijets in the analyses of STAR data.
}

\FullConference{Workshop on Critical Examination of RHIC Paradigms\\
		April 14-17, 2010\\
		Austin, Texas, U.S.A.}

\begin{document}

\section{Introduction}
In this workshop various approaches to the study of nuclear collisions have been presented and compared.  I have been asked to present the approach that emphasizes parton recombination.  Although hadronization by recombination is a process that occurs at late times when the medium density is substantially reduced from the early time, the problem of determining what partons recombine opens up the study to include the physics at all times.  Indeed, while the media created in leptonic, hadronic and nuclear collisions may differ drastically initially, the hadronization processes in the final stages of their evolutions are essentially universal.  Of course, fragmentation is uniquely suitable for leptonic processes, but that is because of the absence of neighboring medium partons.  In a more general approach to hadronization, fragmentation is a subclass of recombination.  Thus the issue returns to the question of what recombines, and the partonic system that is formed by a collision is the real object of investigation.

At the suggestion of the organizers my talk included topics on the formulation of recombination in the beginning many years ago.  The valon model was introduced as a link between deep inelastic scattering at high $Q^2$ and hadronic wave function at low $Q^2$, which in turn made possible a description of the recombination function in the time-reversed process.  Application was then made to the low-$p_T$ process of $p+p \rightarrow \pi + X$.  Shower parton distributions were derived from the fragmentation functions (FF) treated as the result of recombination of shower partons.  Such an approach led to the possibility of relating the FFs of partons into mesons $(D^M)$ and baryons $(D^B)$, thus closing a conceptual gap in the relationship of the FFs among various species of hadrons produced.

In nuclear collisions one usually identifies three domains of $p_T$ where different approaches are separately most suitable:  low region $(p_T < 2$ GeV/c) where hydrodynamics has been successful, high region $(p_T > 6$ GeV/c) where lower-order calculations in pQCD is valid, and intermediate region $(2 < p_T < 6$ GeV/c) where recombination dominates over fragmentation.  Since the ratio of $D^P/D^\pi$ is small, it has been hard for the fragmentation paradigm to reproduce the large $p/\pi$ ratio at intermediate $p_T$ in nuclear collisions.  Even in $pA$ collisions the Cronin effect is larger for produced proton than for produced pion and therefore puts considerable difficulty in the conventional interpretation that the effect is due only to initial-state transverse broadening followed by fragmentation.  Recombination can reproduce the data quite readily.  Space does not permit me to review here all those topics.  A summary of them with extensive references to the original articles can be found in \cite{rh}.  In the text below I discuss the more recent development in nuclear collisions.  I chose two topics of particular relevance to this Workshop because they depart from the conventional wisdom.  One is on azimuthal dependence at low $p_T$ where the usual explanation is that the variation of pressure gradient on $\phi$ leads to elliptic flow.  We consider the effects of semihard scatterings that precede the completion of thermalization?  The other topic is two-jet recombination at LHC that can overwhelm fragmentation for hadrons formed at $p_T$ around 10 GeV/c.

At the end I offer some thoughts on possible common ground with the 2-component model of the UW-UTA alliance in their interpretation of the inclusive distribution and auto-correlation determined in their analyses of the data from heavy-ion collisions at RHIC \cite{tt}.

\section{Azimuthal Dependence at Low $p_T$}
It is conventionally thought that for $p_T < 2$ GeV/c only soft physics is involved and hydrodynamics provides a satisfactory description of the elliptical flow coefficient $v_2(p_T)$ \cite{kh, dat}.  However, the theory assumes rapid thermalization, achieved at $\tau_0=0.6$ fm/c.  Whether that can be proven is a separate issue.  What is relevant to the following discussion is the question on whether semihard scattering with parton $k_T$ less than 3 GeV/c can have any effect that supersedes the hydro result, since there are many such scatterings and they take place in $\tau^ <_\sim 0.1$ fm/c.  The point is that whereas hard scatterings are rare and can be ignored in the treatment of the bulk, there may come a point when the transverse momenta $k_T$ are low enough such that semihard partons are pervasive and not negligible, yet not accounted for by the hydro treatment of a macroscopic medium in local thermal equilibrium.  A possible empirical basis upon which such questions on the "conventional wisdom" can be raised is the implication of the existence of minijets by the two-component analysis of the STAR data \cite{tt, tat}.  Without necessarily endorsing the specifics of the two-component model, we examine the possibility that semihard partons emitted from the surface can lead to the observed azimuthal anisotropy, even if thermlization cannot be completed until after $1$ fm/c.

Our focus in this section is on ridge production at midrapidity, given the experimental fact that the extraction of the $\phi$ dependence of associated particles relies on a section of $\Delta\eta > 0$ that characterizes the ridge (elongation in $\Delta\eta$) \cite{jp}.  Ridge formation at large $\Delta\eta$ is a separate problem and is to be addressed separately \cite{ch}.  Here our emphasis is in the study of $\phi$ dependence in the single-particle distribution $\rho_1(p_T, \phi, b)$ at $\eta \approx 0$, arising from semihard partons without any triggers.  Our point is that if semihard partons are copiously produced near the surface with a $\phi$ anisotropy that is calculable from geometry, then they can have nontrivial consequences on $\rho_1(p_T, \phi, b)$ in a way similar to how ridge structure can develop in the particle distribution associated with a high-$p_T$ trigger \cite{rch, chy, hz}.  Although the use of ridge in the nomenclature referring to such structure was initiated by Putschke \cite{jp}, the discovery of the same behavior (narrow in $\Delta\phi$ but broad in $\Delta\eta)$ occurred earlier in autocorrelation without triggers \cite{ja}.

When a semihard scattering occurs deep in the interior of the nuclear overlap, the scattered partons get absorbed by the medium and become a part of the bulk.  If it occurs near the surface, one of them can get out after losing some energy, which in turn can enhance the thermal energy of the partons near its trajectory.  Recombination of the enhanced thermal partons gives rise to the ridge particles locally.  Since there can be many semihard scatterings in each event of central or mid-central collision, the sum of all such ridge particles produced gives rise to a $\phi$ dependence in $\rho_1(p_T, \phi, b)$.  Our assumption is that the base upon which the ridge sits is $\phi$ independent.  Thus we write
\begin{eqnarray}
\rho_1(p_T, \phi, b) = B(p_T, b) + R(p_T, \phi, b). \label{1}
\end{eqnarray}
The first term on the right side is not the hydro bulk, which has $\phi$ dependence and is all of $\rho_1(p_T, \phi, b)$ at $p_T < 2$ GeV/c.  We call it Base, which is $\phi$ independent, on the grounds that the anisotropic geometry of the overlap at $b >0$ leads to calculable asymmetry of the semihard partons near the surface at early time and produces the only $\phi$ dependence of $\rho_1(p_T, \phi, b)$ at late time through the ridge term $R(p_T, \phi, b).$  In that picture it is necessary to demonstrate that the ridge structure in the single-particle distribution can be related to the ridge yield in two-particle correlation with a detected high-$p_T$ trigger, since the same physics is involved.  In that way we have a unified picture without patching together different pieces of physics to distinguish different kinds of ridges.

For notational simplicity, let us use $\rho_2^R(\phi_1, \phi_2)$ to denote the ridge part of the two-particle distribution with a hard or semihard parton at $\phi_1$ and several thermal partons at $\phi_2$ that can form the detected hadron (pion or proton) with the same $\phi_2$.  The last term in (\ref{1}), abbreviated as $R(\phi_2)$, gives the ridge distribution whether or not the parton at $\phi_1$ leads to a trigger.  Thus we have
\begin{eqnarray}
R(\phi_2) \propto \int d\phi_1\rho_2^R(\phi_1, \phi_2). \label{2}
\end{eqnarray}
On the other hand, if events are selected by trigger particles at $\phi_1$, the ridge yield is integrated over all associated particles at $\phi_2$, so we should have for the yield
\begin{eqnarray}
Y^R(\phi_1) \propto \int d\phi_2\rho_2^R(\phi_1,\phi_2). \label{3}
\end{eqnarray}
Our aim is to give a theoretical description that quantifies the above relations as well as to show that they are consistent with the data at hand \cite{hz}.

The basic premise of our approach is that the elliptical nuclear overlap of the initial configuration in the transverse plane at $y \approx 0$ is sufficient to describe the $\phi$ dependence of the ridge distribution without carrying out a detailed hydrodynamical calculation.  Our assumption is that on average the final hadrons are directed normal to the surface of the initial ellipse.  Thus for any point on that surface with its normal vector denoted by $\phi_2$, one can ask what semihard partons at $\phi_1$ can contribute to the formation of a ridge particle at $\phi_2$.  This question was addressed earlier in an article on the correlated emission model (CEM) \cite{ch2}, where a Gaussian width of $\sigma \approx 0.33$ was found to give a good fit of the $\phi_1$ dependence of the ridge yield \cite{af}.  That is, $\Delta\phi = \phi_2 - \phi_1$ cannot be too large if there is to be coherence of the enhanced thermal partons to form a detectable ridge particle.  If $\Delta\phi$ is much larger than $\sigma$, then the effect of the energy loss of the semihard parton is spread out over a large angular range due to the radial flow of the thermal partons, too diffused to result in effective coalescence.  The physics can be simply summarized by a line segment $S(\phi_2, b)$ on the surface around $\phi_2$ through which semihard partons can contribute .  For every $b$, $S(\phi, b)$ can be analytically determined in terms of elliptic integrals of the second kind \cite{hz}, the details of which need not be reproduced here.  Its dependences on $b$ for $\phi = 0$ and $\pi/2$ are shown in Fig.\ 1.  For mid-central collision ($b \approx 1$ in units of $R_A$) where the initial ellipse is neither too narrow nor too broad, there is a large difference between emission at $\phi \approx 0$ and that at $\phi \approx \pi/2$.  That sounds like the result of hydro calculation, but no hydrodynamics has been applied.

\begin{figure} 
\hspace{2.5cm}
\includegraphics[width=.6\textwidth]{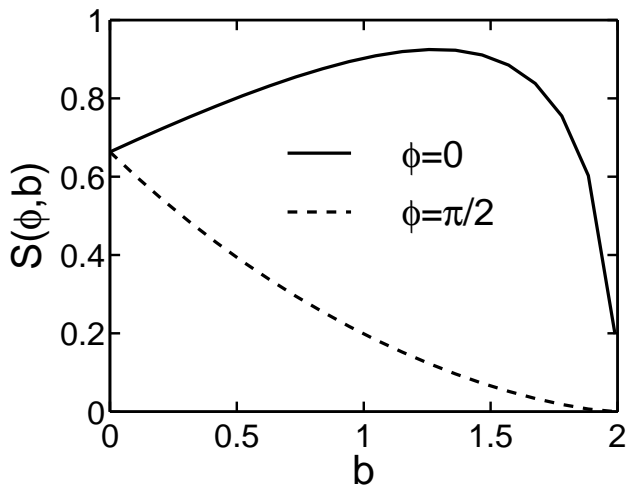} 
\caption{Surface segment $S(\phi,b)$ vs normalized impact parameter $b$ in units of $R_A$ for $\phi=0$ (solid line) and $\phi=\pi/2$ (dashed line).} \label{fig1} 
\end{figure}

\begin{figure} 
\hspace{2.5cm}
\includegraphics[width=.6\textwidth]{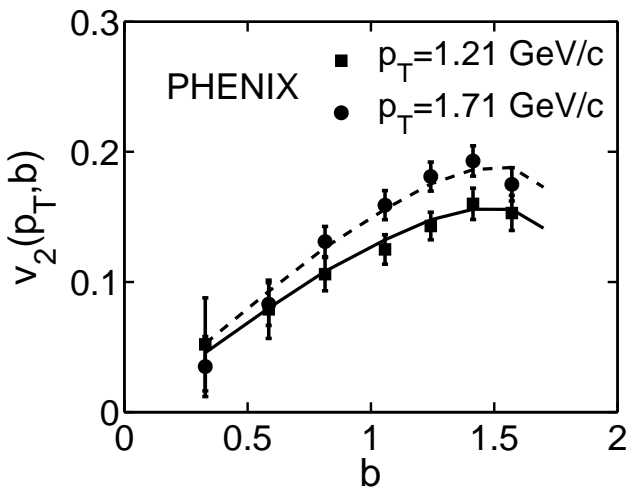} 
\caption{Elliptic flow coefficient $v_2(p_T,b)$ vs $b$ for $p_T=1.21$ GeV/c (solid) and 1.71 GeV/c (dashed). Data are from \cite{sa}.} \label{fig2} 
\end{figure}

Since the probability of producing the ridge particles depends on the density of semihard partons that give rise to the enhanced thermal partons, we thus have
\begin{eqnarray}
R(p_T, \phi, b) \propto S(\phi, b). \label{4}
\end{eqnarray}
The proportionality factor involves the nuclear density and an exponential dependence on $p_T$.  Since $S(\phi, b)$ prescribes the only $\phi$ dependence of $\rho_1(p_T, \phi, b)$, we can determine the strength of the ridge, relative to the base, by fitting $v_2(p_T, b)$ \cite{hz}.  That is shown in Fig.\ 2.  The ridge yield, on account of (\ref{3}), also depends on the surface segment according as
\begin{eqnarray}
Y^R(\phi_1) \propto S(\phi_1, b). \label{5}
\end{eqnarray}
Fig.\ 3 shows how well the data are reproduced by (\ref{5}) with normalization adjusted to render a good overall fit of the 0-5\% data [Fig.\ 2(a)].  Both the normalization of the 20-60\% data [Fig.\ 2(b)] and the shapes for both centralities are well reproduced by the calculated results without further adjustment.  This demonstrates that the triggered and untriggered ridges are related by the same surface segment $S(\phi, b)$, without requiring rapid thermalization or pressure gradients.

\begin{figure} 
\hspace{2.5cm}
\includegraphics[width=.6\textwidth]{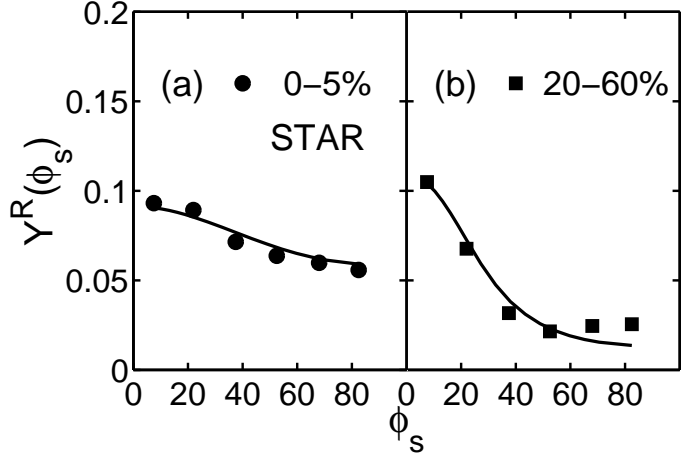} 
\caption{Ridge yield per trigger vs trigger angle $\phi_s$ relative to reaction plane for (a) 0-5\% and (b) 20-60\% centralities. Solid lines are calculated results given in \cite{hz} based on (2.5), and the data are from \cite{af}.} \label{fig3} 
\end{figure}

\section{Scaling Behavior at High $p_T$}
At high $p_T$ it is necessary to take into account hard partons traversing the nuclear medium.  PHENIX has data on $R_{AA}$ that are separated into bins of $p_T, \phi$ and $N_{\rm part}$ \cite{sa}.  The dependencies on the three variables are so complicated that there is a need for organizational simplification.  Obviously, $\phi$ and $N_{\rm part}$ (or $b$, as we prefer to use) are related by geometry.  We find that an average dynamical path length can be defined, in terms of which $R_{AA}$ has universal scaling behavior that can in turn be reproduced in our theoretical calculation \cite{hy}.

Consider a collision at $b$ with a high-$p_T$ particle detected at $\phi$.  Suppose that a hard parton is produced at the position $x_0, y_0$ in the initial elliptical overlap with a transverse momentum $k$ directed at $\phi$.  We consider only the transverse plane at mid-rapidity.  Let us define a geometrical path length by 
\begin{eqnarray}
\ell(x_0, y_0, \phi, b) = \int_0^{t_1(x_0, y_0, \phi, b)} dt\, D[x(t), y(t)], \label 6
\end{eqnarray}
where $t$ is just a label to mark the trajectory from the point of origination to the exit point $t_1$.  $D$ is the density of the medium that the hard parton traverses, and can be calculated from the almond-shaped overlap region.  Clearly, if the nuclear medium has hypothetically low density, it is sensible to regard $\ell$ as zero even though $t_1\ne 0$.  We do not consider medium expansion for two reasons:  (a) if we did, it would be necessary to involve hydrodynamics that is an additional theoretical input we want to avoid, and (b) an expansion would lower $D$ but lengthens $t$ with compensating effects.  There are subtleties in mapping the almond to the elliptical boundaries, especially when $(x_0, y_0)$ is not too far from the surface \cite{hy}.  For long transit time, which is less important, longitudinal expansion should be taken into consideration, and we rely on the dynamical scaling behavior found in \cite{sw} to regard our procedure as being insensitive to that expansion.  Basically (\ref{6}) is precise and essentially geometrical.

To take the dynamics of momentum degradation into account, we define the dynamical path length 
\begin{eqnarray}
\xi = \gamma \ell(x_0, y_0, \phi, b) \label{7}
\end{eqnarray}
with $\gamma$ to be determined, and the average is 
\begin{eqnarray}
\bar\xi(\phi,b)=\gamma\int dx_0dy_0 \ell(x_0,y_0,\phi,b) Q(x_0,y_0,b), \label{8}
\end{eqnarray}
where $Q(x_0, y_0, b)$ is the probability of producing a hard parton at $x_0, y_0$ and can be determined from the thickness functions of the colliding nuclei at $b$.  The significance of considering $\bar\xi(\phi, b)$ is that for every pair of $(\phi, b)$, one can calculate $\bar\xi$ and therefore $R_{AA}(\bar\xi)$, given the data point at $(\phi, b)$.  What we have found is that for every bin of $p_T$ the data on $R_{AA}$ exhibit universal behavior in $\bar\xi$, independent of the variations of $(\phi, b)$ for each $\bar\xi$.  That is shown in Fig.\ 4, which is a property of the data independent of theoretical input.  Only two bins of $p_T$ are shown, but they exemplify the dependence on $\bar\xi$ for 10 bins in the interval $2 < p_T < 10$ GeV/c.  A specific value of $\gamma$ has been used in the plots, but that should not affect the existence of the scaling behavior, since $\gamma$ only determines the horizontal scale.  Obviously, the scaling behavior provides considerable economy in the presentation of the data, and makes possible an efficient comparison with the results of theoretical calculation.

\begin{figure} 
\hspace{-.7cm}
\includegraphics[width=1.1\textwidth]{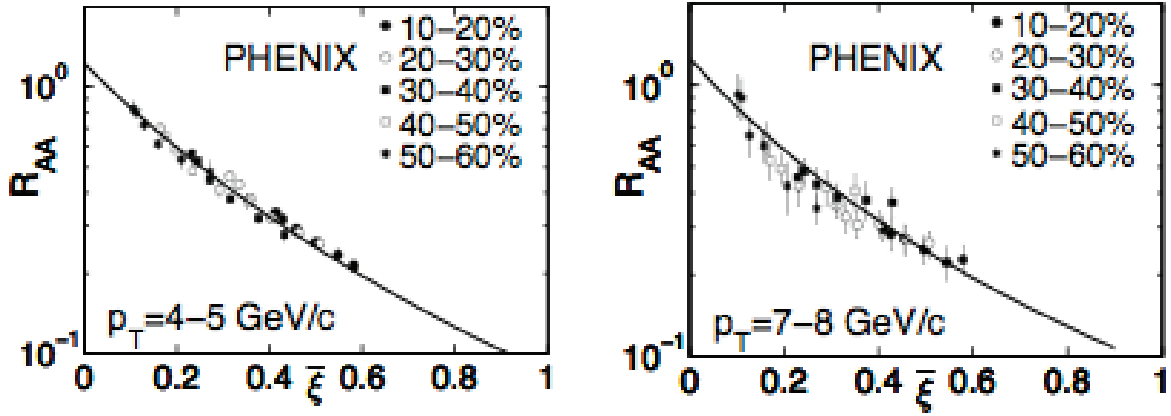} 
\caption{Experimental data on $R_{AA}$ for $p_T$=4-5 (left) and 7-8 GeV/c (right) from Ref.\ \cite{sa} are plotted against the scaling variable $\bar\xi$. The $\phi$ values are defined by $\phi=n\pi/24$ with $n$ being 1,3,.. from left to right, all having the same symbol for each centrality. The solid lines are the same as the solid lines in Fig.\ 5 that represent the  results from theoretical calculation.} \label{fig4} 
\end{figure}

For $p_T > 2$ GeV/c it is necessary to consider the shower partons, which in turn depend on the hard parton distribution $F_i(q,\xi)$ at the surface with momentum $q$.  It is related to the parton momentum $k$ at the point of creation by
\begin{eqnarray}
F_i(q,\xi)=\int dkkf_i(k)G(k,q,\xi), \label{9}
\end{eqnarray}
where $f_i(k)$ is the $k$ distribution of parton type $i$, and $G(k,q,\xi)$ is the degradation function
\begin{eqnarray}
G(k,q,\xi)=q\delta(q-ke^{-\xi}). \label{10}
\end{eqnarray}
The $\phi$ dependence is completely imbedded in $\xi$.  In the usual recombination formalism the shower parton distribution can be determined from $F_i(q,\xi)$.  For the thermal partons the surface segment function $S(\phi,b)$ in Fig.\ 1 is needed to incorporate the $\phi$ dependence.  With those ingredients the TS and SS recombination components can be calculated for the single-particle distribution $\rho_1(p_T,\xi)$ and then for $R_{AA}(\bar\xi)$ \cite{hy}.  The results are shown as points in Fig.\ 5 for six values of centrality (denoted by $c$) and six values of $\phi$ (defined by $n\pi/24)$.  All 36 points lie on a universal curve for each bin of $p_T$, when plotted against $\bar\xi$.  The value of $\gamma$ is adjusted to fit $R_{AA}(p_T,\phi,N_{\rm part})$ for $4 < p_T < 5$ GeV/c at various values of $\phi$ and $N_{\rm part}$; it is found to be 
\begin{eqnarray}
\gamma=0.11.   \label{11}
\end{eqnarray}
Thus one set of data has been used to fix the degree of momentum degradation in the nuclear medium; all other results are obtained without further adjustment.

\begin{figure} 
\hspace{-.3cm}
\includegraphics[width=1.05\textwidth]{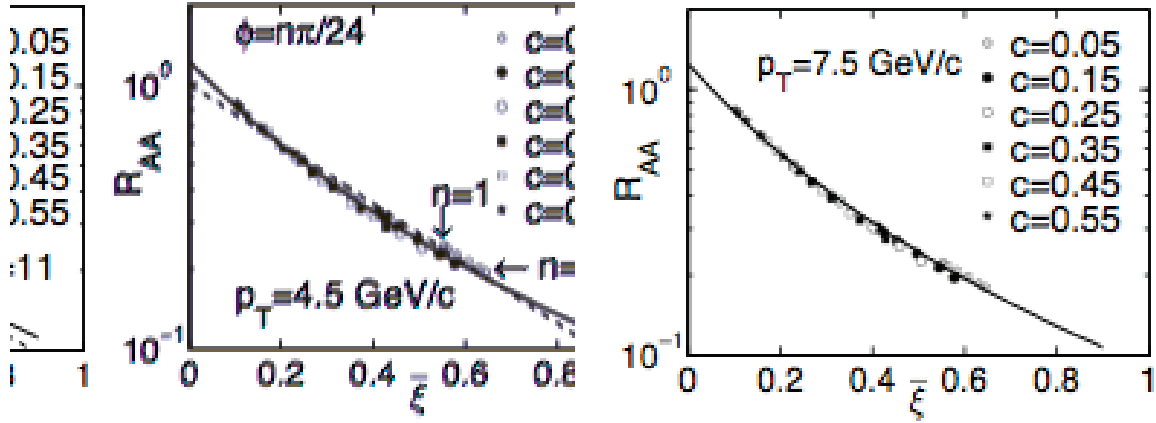} 
\caption{Theoretical result showing scaling behavior of $R_{AA}$ at $p_T=4.5$ GeV/c (left) and $p_T=7.5$ GeV/c (right) for six values of centrality. The points for different $\phi$ values between 0 and $\pi/2$ have the same symbol when the centralities are the same,  with $\phi$ increasing from left to right with $n=1,3,\cdots,11$. The solid lines are the best fits; the dashed line is an exponential fit by (3.7).} 
\label{fig5} 
\end{figure}

The solid lines in Fig.\ 5 are analytic curves to represent the calculated points.  Those curves are transported to Fig.\ 4 to facilitate the comparison between theory and experiment.  The agreement is evidently very good.  Thus the recombination model satisfactorily reproduces all the data points for all $p_T > 2$ GeV/c and all $\phi$ and $c$ with one adjustable parameter $\gamma$, whose implication on momentum degradation can be  more simply described by the dashed line in the left panel of Fig.\ 5.  It is a straight-line fit of $R_{AA}(\bar\xi)$, giving
\begin{eqnarray}
\left. R_{AA}^{\pi}(\bar\xi)\right|_{p_T=4.5}=\exp(-2.6\bar\xi), \label{12}
\end{eqnarray}
describing an exponential suppression at increasing $\bar\xi$, which is physically reasonable, and is dependent only on the average path length, independent of separate values of $\phi$ and $b$, which is not surprising, yet nevertheless a new finding.  It is hard to relate the numerical value of $\gamma$ to other physical quantities that we are familiar with.

In extending our consideration to LHC we find that with all the physical mechanisms remaining the same, the scaling feature in $\bar\xi$ is preserved, as shown in Fig.\ 6(a) in a linear plot for $p_T=10$ GeV/c.  However, the hadronization process need not be the same.  At LHC the density of hard partons created is so high that the shower partons from neighboring jets can recombine to form hadrons.  We do not mean large $x_T$, for which parton density is low whether at RHIC or LHC.  At fixed $p_T$ the parton density increases with $\sqrt{s}$, so it is possible that at some energy the shower cones of adjacent jets may overlap.  In that case the recombination of shower partons arising from parallel jets can produce hadrons copiously, since the hard-parton momenta, $k_1$ and $k_2$, can be less than $p_T$, as opposed to being greater than $p_T$ for fragmentation.  In Fig.\ 6(b) we show the result for $p_T=10$ GeV/c at LHC for a particular choice of the overlap parameter $\Gamma$ in two-jet recombination \cite{hy}.  There are two striking features in that figure.  One is that the scaling behavior in Fig.\ 6(a) is badly broken.  The other is that the magnitude of $R_{AA}$ is extraordinarily large, with the possibility of exceeding 1.  The observation of either feature would be a clear signature of significant departure from RHIC physics, the latter would stand out distinctly from other predictions.  The origin of scaling violation and large $R_{AA}$ is in the definiton of $R_{AA}$, where the normalization factor involves $N_{\rm coll}$ for single jets.  When 2-jet recombination becomes important, normalizing $R_{AA}$ by $N_{\rm coll}^2$ will restore $\bar\xi$ scaling and lower $R_{AA}$.  We add that $R_{AA}$ for $p$ production at LHC can be even larger due to the recombination of three shower partons \cite{hy1}.

\begin{figure} 
\hspace{-.6cm}
\includegraphics[width=1.1\textwidth]{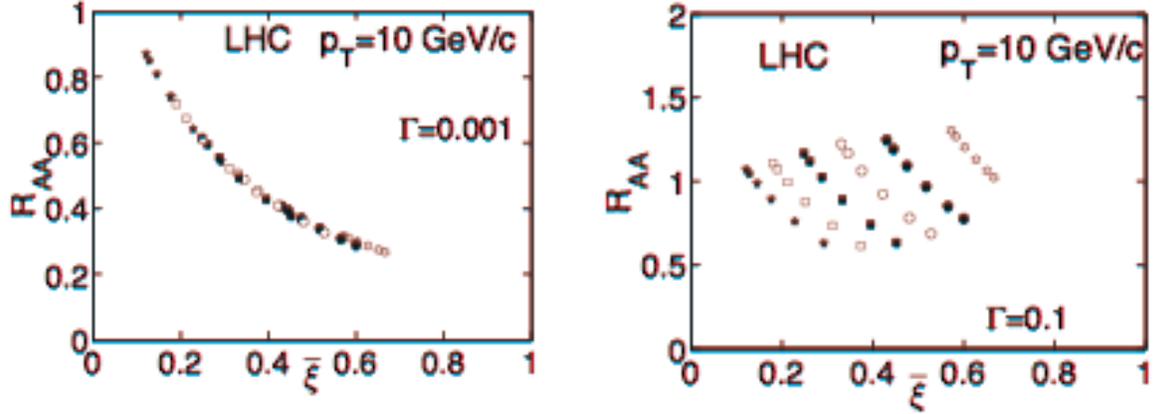} 
\caption{(Left) Scaling behavior  of $R_{AA}$ at LHC for $p_T$=10 GeV/c and negligible 2-jet recombination. The symbols are the same as in Fig.\ 5 for various values of $c$ and $\phi$. 
(Right) Non-scaling behavior  of $R_{AA}$ at LHC for $p_T$=10 GeV/c when 2-jet recombination is dominant.} 
\label{fig6} 
\end{figure}

It should be pointed out that if the features in Fig.\ 6(b) are not found at LHC, it does not imply the failure of the recombination model.  It just means that the overlap of neighboring jet cones is not of high probability to enhance 2-jet recombination.  The observation of any features between Fig.\ 6(a) and (b) would provide some empirical ground for quantifying the recombination function of adjacent jets.

\section{Comparative Comments}
Since the aim of this Workshop is to exchange ideas on different approaches to RHIC physics, it seems appropriate to end this talk with some comments on the similarities and differences among some of the models.  In particular, there seems to exist some common ground between the point of view advanced in Sec.\ 2 and the two-component model advocated by the UW-UTA alliance \cite{tt}.  That model consists of a soft $S$ and a hard $H$ component with the most notable observation that in Au-Au collisions $H$ has a strong enhancement at small transverse rapidity $y_t$.  It is interpreted as being due to minijets which peak at around  $y_t = 2.7$, corresponding to $p_T \approx 1$ GeV/c.  Putting aside any reservation about the procedure of separating the two components, the observation of a pronounced peak in the $y_t$-$y_t$ correlation and the necessity of including a 2D Gaussian peak in order to fit the autocorrelation data on $\eta_\Delta$-$\phi_\Delta$ in heavy-ion collisions provide some empirical evidence for the existence of minijets \cite{tt}.  It is claimed that the hadron spectra of the hard component reveal the dominance of parton fragmentation all the way down to below $p_T \sim 1$ GeV/c, and that the abundance of surviving minijets conflicts with near-ideal hydrodynamics \cite{tat}.

In Sec.\ 2 we have considered the possible effect of semihard partons on $\phi$ anisotropy before thermal equilibrium is established.  Since semihard scattering is not rare for $k_T < 3$ GeV/c, there is a possibility that there may be a correspondence between those semihard partons and minijets.  The hadronization mechanisms may be quite different, ours being the recombination of enhanced thermal partons, while the 2-component model stresses fragmentation.  Our ridge component in the single-particle distribution contains the $\phi$ dependence, corresponding to the hard component that has the low-$p_T$ enhancement.  Autocorrelation exhibits $\phi_\Delta$ dependence after integation over $\phi_\Sigma$.  It would be of interest to know the dependence on $\phi_\Sigma$ before integration (admittedly a bit hard to achieve), since that would give a possible hint of the $\phi$ dependence of $S(\phi,b)$, the boundaries of which are shown in Fig.\ 1, but it is totally known analytically \cite{hz}.  The ridge yield has the curious property that, when the trigger is along the direction of the reaction plane $(\phi_s=0)$, the yield increases as the collision becomes less central.  That is a property shown by the STAR data in Fig.\ 3 and corresponds to the solid line in Fig.\ 1.  Since $S(\phi,b)$ is the surface segment for semihard partons whether they lead to trigger [cf.\ (\ref {5})] or not [cf.\ (\ref {4})], it would be good to verify its relevance to both the triggered events \cite{af} and the untriggered autocorrelation.

There are numerous questions in my mind about the 2-component model when the collision energy is raised to those at LHC.  As the semihard parton density becomes high, the recombination of adjacent jets will at some point become unavoidable.  The separation of soft and hard components will also become more difficult, especially if the hard component is so identified with fragmentation, as done at RHIC \cite{tat}.  But that will be a problem for all theoretical models that exclude recombination.  Of course, recombination is only one of many paradigms that has not yet been proved to be universal.  I encourage the organizers to reconvene this Workshop in two or three years and review whether the RHIC paradigms will have significantly changed at LHC.

\section*{Acknowledgment}
It has been a great pleasure to participate in this Workshop, for which I am very grateful to Lanny Ray and Tom Trainor for inviting me.  Most of the work reported here was done in collaboration with C. Chiu, C. B. Yang and L. Zhu. This work was supported, in part,  by the U.\ S.\  Department of Energy under Grant No. DE-FG02-96ER40972.


\end{document}